\documentclass{emulateapj}
\usepackage{amsmath}     
\usepackage{wasysym}           
\usepackage{graphicx}
\usepackage{amssymb}
\usepackage{epstopdf}
\usepackage{mathrsfs}
\usepackage[T1]{fontenc}
\usepackage{natbib}

\begin{document}
\title{Viscous boundary layers of radiation-dominated, relativistic jets. II. The free-streaming jet model}
\author{Eric R. Coughlin\altaffilmark{1} and Mitchell C. Begelman\altaffilmark{1}}
\affil{JILA, University of Colorado and National Institute of Standards and Technology, 440 UCB, Boulder, CO 80309}
\email{eric.coughlin@colorado.edu, mitch@jila.colorado.edu}
\altaffiltext{1}{Department of Astrophysical and Planetary Sciences, University of Colorado, UCB 391, Boulder, CO 80309}

\begin{abstract}
We analyze the interaction of a radiation-dominated jet and its surroundings using the equations of radiation hydrodynamics in the viscous limit. In a previous paper we considered the two-stream scenario, which treats the jet and its surroundings as distinct media interacting through radiation viscous forces. Here we present an alternative boundary layer model, known as the free-streaming jet model -- where a narrow stream of fluid is injected into a static medium -- and present solutions where the flow is ultrarelativistic and the boundary layer is dominated by radiation. It is shown that these jets entrain material from their surroundings and that their cores have a lower density of scatterers and a harder spectrum of photons, leading to observational consequences for lines of sight that look ``down the barrel of the jet.'' These jetted outflow models may be applicable to the jets produced during long gamma-ray bursts and super-Eddington phases of tidal disruption events.
\end{abstract}


\section{Introduction}
Particles, magnetic fields, and radiation all contribute to the propulsion of relativistic jets, though the relative contribution of each is still an open matter of debate. In certain situations, however, the mechanism responsible for launching the jet operates simultaneously with the release of a large amount of energy in the form of radiation, making it likely that photons dominate the bulk energetics. This scenario occurs, for example, during the super-Eddington phase of jetted tidal disruption events (TDEs), such as \emph{Swift} J1644+57 \citep{bur11, zau11} and \emph{Swift} J2058+05 \citep{cen12}. Radiation-dominated jets should also be present in the collapsar picture of long gamma-ray bursts (GRBs; \citealt{ree92, woo93, mes93, mac99, pir04}), where the energy released in the form of gamma-rays is ultimately derived from accretion onto a black hole, the associated accretion luminosity exceeding the Eddington limit of the hole by more than ten orders of magnitude. In both of these cases, the propagation of the radiation-dominated jet is modulated by the presence of a radiation pressure-supported environment; for super-Eddington TDEs, this environment is in the form of a highly inflated, quasi-spherical torus of fallback debris \citep{cou14a}, while a ``cocoon'' of shocked jet material \citep{mor07, lop13} and the overlying stellar envelope itself \citep{mat03, woo06} serve as the confining medium for GRBs.

In a companion paper (\citealt{cou15}, hereafter Paper I), we presented a model that describes the viscous interaction of a radiation-dominated, relativistic jet with its surrounding medium. In that analysis we treated the jet and its surroundings as two separate fluids, interacting with one another via small anisotropies in the comoving radiation field that are explicitly accounted for in the equations of radiation hydrodynamics in the viscous limit. This two-stream approximation, in agreement with the non-relativistic analysis of \citet{ara92}, demonstrates the manner in which the shear between the two fluids carves out a region of low density material within the boundary layer between them. We also deduced the dependence of the boundary layer thickness on the asymptotic properties of the jet and the ambient medium. 

These models also show, however, that the contact discontinuity separating the jet and its surroundings, necessary for maintaining their respective identities, results in the likely unphysical vanishing of the density of scatterers along that surface of separation. The contact discontinuity also prevents the jet from entraining ambient material; since the jet in the two-stream model we considered had an infinite amount of momentum, any entrainment or lack thereof is formally inconsequential to the evolution of the system. However, realistic jets -- those with finite extent -- will almost certainly engulf more material as they expand into their surroundings; because the total amount of momentum in the system, which is realistically finite, must be conserved, that entrainment will then result in an overall deceleration of the outflow that cannot be captured in the two-stream model.

In view of these unphysical properties of the two-stream treatment -- the vanishing of the mass density of scatterers along the contact discontinuity and the lack of entrainment -- we present here an alternative boundary layer scenario to describe the interaction of a relativistic, radiation-dominated jet with its surroundings. This ``free-streaming jet'' model, which has a well-known counterpart in the non-relativistic, incompressible limit {}{(see Chapter 10, Section 12 of} \citealt{kun08}{}{)}, assumes that the jet is injected through a narrow opening into a static, homogenous medium, and that far enough from that opening the entire system can be modeled as a single fluid that is independent of the details at the injection point. By considering the jet and the ambient medium as one fluid we obviate the need for a contact discontinuity, which we demonstrate allows the density of scatterers to remain finite throughout the flow and for the jet to entrain material.

In section 2 of this paper we present the equations of radiation hydrodynamics in the viscous limit. Section 3 uses those equations to analyze the free-streaming jet model, and we demonstrate the existence of approximate self-similar solutions in the limit that the interaction between the jet and the ambient medium is concentrated in a thin boundary layer. In section 4 we discuss the implications of our model and make comparisons to the two-stream scenario, and in section 5 we conclude and consider the application of the free-streaming jet model to super-Eddington TDEs, GRBs{}{, and other astronomical sources}.

\section{Governing equations}
When changes in fluid quantities over the mean free path of a photon are small, radiation behaves like an effective viscosity and transfers momentum and energy between neighboring fluid elements. The precise form of the viscosity can be determined by investigating the general relativistic Boltzmann equation, which was recently done by \citet{cou14b} for the case where Thomson scattering dominates the interactions between the photons and scatterers in the fluid rest frame. In this limit, they found that the relativistic equations of radiation hydrodynamics for a cold gas (gas pressure much less than the gas rest mass density and radiation pressure) are ({}{see their equation (49);} see also Paper I):

\begin{multline}
\nabla_{\mu}\bigg{[}\bigg{\{}\rho'+\frac{4}{3}e'\bigg{(}1-\frac{10}{9}\frac{1}{\rho'\kappa}\nabla_{\alpha}U^{\alpha}\bigg{)}\bigg{\}}U^{\mu}U^{\nu}\bigg{]}+\frac{1}{3}g^{\mu\nu}\partial_{\mu}e' \\ 
-\frac{8}{27}\nabla_{\mu}\bigg{[}\frac{e'}{\rho'\kappa}\Pi^{\mu\sigma}\Pi^{\nu\beta}\bigg{(}\nabla_{\sigma}U_{\beta}+\nabla_{\beta}U_{\sigma}+g_{\beta\sigma}\nabla_{\alpha}U^{\alpha}\bigg{)}\bigg{]} \\ 
-\frac{1}{3}\nabla_{\mu}\bigg{[}\frac{e'}{\rho'\kappa}\bigg{(}\Pi^{\mu\sigma}U^{\nu}+\Pi^{\nu\sigma}U^{\mu}\bigg{)}\bigg{(}4U^{\beta}\nabla_{\beta}U_{\sigma}+\partial_{\sigma}\ln{}e'\bigg{)}\bigg{]} = 0 \label{radhydroco}.
\end{multline}
Here the speed of light has been set to one, Greek indices range from 0 -- 3, $\rho'$ is the fluid rest frame mass density of scatterers, $e'$ is the fluid rest frame radiation energy density, $\kappa$ is the scattering opacity (in units of cm$^2$ g$^{-1}$), $g_{\mu\nu}$ is the metric of the spacetime, $\nabla_{\mu}$ is the covariant derivative, $U^{\mu}$ is the four-velocity of the flow, and $\Pi^{\mu\nu} = U^{\mu}U^{\nu}+g^{\mu\nu}$ is the projection tensor. The Einstein summation convention has been adopted here, meaning that repeated upper and lower indices imply summation. This equation also shows, in agreement with previous findings \citep{bla85, loe92}, that the coefficient of dynamic viscosity, $\eta$, for an optically-thick, radiation-dominated gas is

\begin{equation}
\eta = \frac{8}{27}\frac{e'}{\rho'\kappa} \label{etaeq}.
\end{equation}
We will also write down the gas energy equation, obtained by contracting equation \eqref{radhydroco} with the four-velocity $U_{\nu}$, which gives {}{(see equation (50) of \citealt{cou14b})}

\begin{multline}
\nabla_{\mu}(e'U^{\mu})+\frac{1}{3}e'\nabla_{\mu}U^{\mu} = \frac{4}{3}\frac{10}{9}\nabla_{\mu}\bigg{[}\frac{e'}{\rho'\kappa}U^{\mu}\nabla_{\alpha}U^{\alpha}\bigg{]} \\ 
+\frac{8}{27}\frac{e'}{\rho'\kappa}\bigg{(}\nabla_{\sigma}U_{\beta}+\nabla_{\beta}U_{\sigma}+g_{\sigma\beta}\nabla_{\alpha}U^{\alpha}\bigg{)}\Pi^{\mu\sigma}\nabla_{\mu}U^{\beta} \\
+\frac{1}{3}\Pi^{\mu\sigma}\nabla_{\mu}\bigg{[}\frac{e'}{\rho'\kappa}\bigg{(}4U^{\beta}\nabla_{\beta}U_{\sigma}+\partial_{\sigma}\ln{e'}\bigg{)}\bigg{]} \\ 
+\frac{1}{3}\frac{e'}{\rho'\kappa}\bigg{(}4U^{\beta}\nabla_{\beta}U_{\sigma}+\partial_{\sigma}\ln{e'}\bigg{)}\bigg{(}2U^{\mu}\nabla_{\mu}U^{\sigma}+U^{\sigma}\nabla_{\mu}U^{\mu}\bigg{)}
\label{gasenergyco}.
\end{multline}
To close the system, we require that the normalization of the four-velocity be upheld and that particle flux be conserved:

\begin{equation}
U_{\mu}U^{\mu} = -1,
\end{equation}
\begin{equation}
\nabla_{\mu}\bigg{[}\rho'U^{\mu}\bigg{]} = 0 \label{masscont}.
\end{equation}
Equations \eqref{radhydroco} and \eqref{gasenergyco} -- \eqref{masscont} constitute six linearly independent equations for the six unknowns $U^{\mu}$, $e'$, and $\rho'$.

In addition to the energy density of the radiation, $e'$, one can also calculate the number density of photons by requiring that the number flux, $F^{\mu}$, be conserved. One can show \citep{cou14b} that the equation $\nabla_{\mu}F^{\mu} = 0$ becomes, in the viscous limit, 

\begin{multline}
\nabla_{\mu}\bigg{[}N'U^{\mu}\bigg{]} \\ 
= \nabla_{\mu}\bigg{[}\frac{1}{\rho'\kappa}\bigg{(}\frac{10}{9}N'U^{\mu}\nabla_{\sigma}U^{\sigma}+N'U^{\alpha}\nabla_{\alpha}U^{\mu}+\frac{1}{3}\Pi^{\mu\sigma}\nabla_{\sigma}N'\bigg{)}\bigg{]} \label{fluxeq},
\end{multline}
where $N'$ is the rest-frame number density of photons. Once we solve the equations of radiation hydrodynamics for the four-velocity of the fluid and the mass density of scatterers, we can solve equation \eqref{fluxeq} for the number flux of photons. 

The goal of the next two sections is to apply equations \eqref{radhydroco} and \eqref{gasenergyco} -- \eqref{fluxeq} to the boundary layers established between fast-moving jets and their ambient media. For a more thorough discussion of the nature of the equations of radiation hydrodynamics in the viscous limit, we refer the reader to \citet{cou14b}.

\section{Jetted boundary layer}
In this section we consider the problem where a narrow stream of material is continuously injected into a plane-parallel, static, ambient medium, known as the free-streaming jet problem. If the Reynolds number of the outflow is high, the transition between the stream of material and the external environment will be confined to a thin layer, permitting the use of a boundary layer approximation. 

The basic setup is similar to that of the two-stream problem (see Paper I), with the motion of the {}{2-D, plane-parallel} injected stream predominantly along the $z$-direction, the majority of the variation along $y$, no variation or velocity in $x$, and the point of injection at $y = z = 0$. Now, however, there is no contact discontinuity between the stream and the ambient environment, meaning that the entire system is considered a single fluid. We therefore have less freedom in prescribing the asymptotic characteristics of the jet and the environment; however, this configuration permits mixing between the two media, allowing the outflow to entrain material, which almost certainly occurs in realistic jets.

We can reduce the complexity of equations \eqref{radhydroco} and \eqref{gasenergyco} by assuming that the interaction between the jet and the ambient medium takes place over a thin boundary layer of thickness $\delta{y}$, thin in the sense that $\delta \equiv \delta{y}/\delta{z}$ is a small number when $\delta{z}$ is a typical length along the jet. In this case, $\delta$ scales in an identical fashion to that derived for the two-stream boundary layer, which can be determined by comparing leading-order terms in the boundary layer thickness to the inviscid terms in the gas energy equation. Setting the left-hand side of equation \eqref{gasenergyco} to $\sim e'\Gamma_jv_j/z$, we find, as in Paper I,

\begin{equation}
\delta^2 \sim \frac{1}{\rho'\,\kappa\,{z}\,\Gamma_j\,v_j} \label{deltaeq}.
\end{equation}
Here $v_j$ and $\Gamma_j = (1-v_j^2)^{-1/2}$ are the jet velocity and Lorentz factor, respectively, measured at some characteristic length along the jet axis $z_0$, and $\rho'$ is a characteristic density of scatterers throughout the outflow. Due to the fact that the boundary layer thickness is the same, the boundary layer equations governing the outflow are identical to those found in paper I, which can be compactly written as

\begin{equation}
\nabla_{\mu}\bigg{[}\bigg{(}\rho'+\frac{4}{3}e'\bigg{)}U^{\mu}U^{\nu}\bigg{]}+\frac{1}{3}g^{\mu\nu}\partial_{\mu}e' = \frac{8}{27}\frac{\partial}{\partial{y}}\bigg{[}\frac{e'}{\rho'\kappa}\frac{\partial{U^{\nu}}}{\partial{y}}\bigg{]} \label{bleqs},
\end{equation}
\begin{equation}
\nabla_{\mu}\bigg{[}\rho'U^{\mu}\bigg{]} = 0.
\end{equation}
In paper I we dealt with the $\nu = y$ and $\nu = z$ components of equation \eqref{bleqs} and its contraction with $U_{\nu}$ -- the gas energy equation \eqref{gasenergyco}. For the free-streaming jet problem, however, it will be more convenient to deal with the $\nu = y$ component of equation \eqref{bleqs}, the gas energy equation, and the contraction of equation \eqref{bleqs} with the projection tensor $\Pi^{z}_{\,\,\nu}$, which, as we will see, is the relativistic, viscous counterpart of the Bernoulli equation. Performing a few manipulations, we find that these equations become, respectively, 

\begin{equation}
\frac{\partial{e'}}{\partial{y}} = 0 \label{ymom1},
\end{equation}
\begin{equation}
\nabla_{\mu}\bigg{[}e'U^{\mu}\bigg{]}+\frac{1}{3}e'\nabla_{\mu}U^{\mu} = \frac{8}{27}\frac{e'}{\rho'\kappa}\bigg{(}\frac{\partial{S}}{\partial{y}}\bigg{)}^2 \label{gasen1},
\end{equation}
\begin{equation}
\bigg{(}\rho'+\frac{4}{3}e'\bigg{)}U^{\mu}\nabla_{\mu}S+\frac{1}{3}\Gamma\frac{d{e'}}{d{z}} = \frac{8}{27}\frac{\partial}{\partial{y}}\bigg{[}\frac{e'}{\rho'\kappa}\frac{\partial{S}}{\partial{y}}\bigg{]} \label{zmom1},
\end{equation}
\begin{equation}
\nabla_{\mu}\bigg{[}\rho'U^{\mu}\bigg{]} = 0 \label{masscont1}.
\end{equation}
where

\begin{equation}
S = \ln\bigg{(}\Gamma{v_z}+\sqrt{1+\Gamma^2{v_z}^2}\bigg{)} = \text{arcsinh}(\Gamma{v_z}) \label{Seq}.
\end{equation}
The first of these shows that the radiation energy density, and consequently the pressure, is constant across the boundary layer. In the inviscid limit, equation \eqref{zmom1} can be transformed to give

\begin{equation}
U^{\mu}\nabla_{\mu}\bigg{[}\Gamma\bigg{(}1+\frac{4}{3}\frac{e'}{\rho'}\bigg{)}\bigg{]} = 0,
\end{equation}
which, as we mentioned, is the relativistic generalization of the Bernoulli equation. We will also assume that the ambient energy density is independent of $z$, i.e., that $e'(z) = e'_0$, with $e'_0$ a constant. 

Because it will be convenient, we will change variables from $y$ to $\tau$, where $\tau$ is given by

\begin{equation}
\tau \equiv \int_0^{y}\rho'(\tilde{y},z)\,\kappa\,d\tilde{y} \label{taueq},
\end{equation}
which is related to the optical depth across the boundary layer as measured from the axis. In terms of this variable, equations \eqref{gasen1} and \eqref{zmom1} become, respectively,

\begin{equation}
\frac{4}{3}e'U^{\mu}\nabla_{\mu}\bigg{(}\frac{1}{\rho'}\bigg{)} = \frac{8}{27}e'\kappa\bigg{(}\frac{\partial{S}}{\partial\tau}\bigg{)}^2 \label{gasen2},
\end{equation}
\begin{equation}
\bigg{(}1+\frac{4}{3}\frac{e'}{\rho'}\bigg{)}U^{\mu}\nabla_{\mu}S = \frac{8}{27}e'\kappa\frac{\partial^2S}{\partial\tau^2} \label{zmom2},
\end{equation}
where we have used the assumption that $e'(z) = e'_0$. It should also be noted that the $y$-component of the convective derivative is now with respect to $\tau$, not $y$, i.e.,

\begin{equation}
U^{\mu}\nabla_{\mu} = U^{z}\frac{\partial}{\partial{z}}+\rho'\kappa{\,}U^{y}\frac{\partial}{\partial\tau}.
\end{equation}
In the ensuing section we will seek self-similar solutions to these equations.

\subsection{Self-similar approximation}
We can immediately solve the continuity equation \eqref{masscont1} by introducing the stream function $\psi$ via

\begin{equation}
\Gamma{v_z} = \frac{\partial\psi}{\partial{\tau}} \label{psi1},
\end{equation}
\begin{equation}
\rho'\kappa\,\Gamma{v_y} = -\frac{\partial\psi}{\partial{z}} \label{psi2},
\end{equation}
where the factor of $\kappa$ ensures that $\psi$ remains dimensionless. In paper I we showed that there exist self-similar solutions for $\psi$, the velocity, the comoving density of scatterers and comoving density of photons in terms of the variable $\alpha = y/\delta{y}$, where $\delta{y} \sim \sqrt{z}$. One difference between the two-stream problem and this type of outflow, however, is that we expect the jet to expand into the ambient medium, entraining material in the process. Therefore, as we look farther along the $z$-direction, the amount of inertia -- predominantly in the form of radiation for the systems that we are considering -- contained in the flow will increase. Owing to the conservation of momentum, the $z$-component of the velocity should thus be a decreasing function of $z$. 

In light of this observation, we will assume that the $z$-component of the four-velocity scales as

\begin{equation}
\Gamma{v_z} = \Gamma_jv_j\bigg{(}\frac{z}{z_0}\bigg{)}^{-m}\frac{df}{d\xi} \equiv \zeta\frac{df}{d\xi} \label{vzsim},
\end{equation}
where $z_0$ is a characteristic length scale in the $z$-direction, $m$ is a positive constant, for brevity we defined $\zeta \equiv \Gamma_jv_j(z/z_0)^{-m}$, and $f$ is a function of the self-similar variable $\xi \simeq \tau/\delta\tau$; $\delta\tau$ is the characteristic boundary layer thickness in terms of $\tau$, which, from equations \eqref{deltaeq} and \eqref{taueq} in the previous subsection, is given by $\delta\tau \simeq \rho'\kappa\,\delta{y}$, or

\begin{equation}
\delta\tau^2 \simeq \frac{\rho'\kappa{z}}{\Gamma_jv_j}.
\end{equation}

We are most interested in the behavior of the properties of the outflow when the velocities are relativistic, as this is the limit that is most applicable to sources of astronomical interest, and in order for the self-similar nature of our solutions to be upheld, the dependence of our self-similar functions on the bulk properties of the outflow, such as the $z$-dependence of the jet Lorentz factor along the axis, should be minimal.  A fully self-similar solution is likely impossible here, as the speed of light plays a role in setting a finite scale factor for our solutions and becomes problematic when we try to connect the ultra- and non-relativistic regions of the outflow. We will show, however, that the non-self-similarity of our solutions only affects a small region of the outflow.

In paper I, we showed that the self-similarity of the comoving density of scatterers, $\rho'$, was approximately satisfied, i.e., we could find solutions with $\rho' \sim g(\xi)$. For the case at hand, then, this assumption implies that $g$ is independent of $\zeta$. This means, however, that the observer-frame density of scatterers, given by $\rho = \Gamma\rho' \simeq \zeta\,{g}$, can be made arbitrarily large (as $g$ is independent of $\zeta$ in the self-similar limit), which is a nonsensical result. 

Motivated by this reasoning, we conclude that the comoving density of scatterers is unlikely to be independent of $\zeta$. On the contrary, a more reasonable approximation is that the \emph{observer}-frame density of scatterers varies self-similarly. Investigating equations \eqref{gasen2} and \eqref{zmom2}, we see that the $\rho'$-dependent quantity that enters both is the combination $1+4e'/(3\rho')$. Our expectation that the observer-frame density varies self-similarly then prompts the assumption

\begin{equation}
1+\frac{4}{3}\frac{e'}{\rho'} = \frac{4}{3}\mu\,\Gamma\,g(\xi) \label{rhosim},
\end{equation}
where $\mu \equiv e'/\rho'_0$, $\rho'_0$ being the density of scatterers in the ambient medium. (Note that we could have simply let $\rho'\Gamma = g(\xi)$, but equation \eqref{rhosim}, which is merely a change of variables from the initial assignment $\rho'\Gamma = g(\xi)$, will allow equations \eqref{gasen2} and \eqref{zmom2} to be written in a more compact form.)

We can determine the value of $m$, which controls how rapidly the flow decelerates due to entrainment, by integrating the momentum flux, $\dot{P} \simeq e'\Gamma^2v_z^2$ for a radiation-dominated system, over the entire boundary layer, and requiring that the result be independent of $z$. This is equivalent to requiring that the total momentum contained in the outflow be conserved. Integrating $e'\Gamma^2v_z^2$ from $y = -\infty$ to $\infty$, using equations \eqref{vzsim} and \eqref{rhosim}, taking the ultrarelativistic limit and changing variables from $y$ to $\xi$, we find 

\begin{equation}
\dot{P} \simeq \bigg{(}\frac{z}{z_0}\bigg{)}^{\frac{1}{2}-2m}\int_{-\infty}^{\infty}\left(\frac{df}{d\xi}\right)^2d\xi.
\end{equation}
Since the integral in this equation is a constant that is greater than zero, we find that $m = 1/4$ if the momentum flux is conserved. 

With this value of $m$, the $z$-component of the four-velocity scales as

\begin{equation}
\Gamma{v_z} = \Gamma_jv_j\bigg{(}\frac{z}{z_0}\bigg{)}^{-1/4}\frac{df}{d\xi} \label{vzsim2},
\end{equation}
and the self-similar variable $\xi$ is given by

\begin{equation}
\xi = \tau\,\sqrt{\frac{9\Gamma_j^2v_j^2}{2\rho'_0\kappa{z_0}}}\bigg{(}\frac{z}{z_0}\bigg{)}^{-3/4} \label{xieq}.
\end{equation}
We {}{then} find from equation \eqref{psi1} that the stream function must satisfy

\begin{equation}
\psi = \sqrt{\frac{2}{9}\rho'_0\kappa{z_0}}\bigg{(}\frac{z}{z_0}\bigg{)}^{1/2}f. 
\end{equation}
Equation \eqref{psi2} then gives the $y$-component of the four-velocity:

\begin{equation}
\rho'\kappa\Gamma{v_y} = -\sqrt{\frac{2}{9}\frac{\rho'_0\kappa}{z_0}}\bigg{[}\frac{1}{2}\bigg{(}\frac{z}{z_0}\bigg{)}^{-1/2}f+z_0\bigg{(}\frac{z}{z_0}\bigg{)}^{1/2}\frac{\partial\xi}{\partial{z}}f_{\xi}\bigg{]} \label{vysim},
\end{equation}
{}{where, both here and in future equations, a subscripted $\xi$ on the functions $f$ and $g$ denotes differentiation with respect to $\xi$ and the number of subscripts indicates the number of derivatives, i.e., $f_{\xi} = df/d\xi$, $f_{\xi\xi} = d^2f/d\xi^2$, etc.}

Inserting equations \eqref{rhosim}, \eqref{vzsim2} and \eqref{vysim} into equations \eqref{gasen2} and \eqref{zmom2}, changing variables from $\tau$ to $\xi$ and performing a bit of algebra, we find that they become, respectively,

\begin{equation}
-\frac{1}{2}\Gamma{f}g_{\xi} = \zeta\bigg{(}\bigg{(}\frac{\partial{S}}{\partial\xi}\bigg{)}^2-v_z\frac{\partial^2{S}}{\partial\xi^2}\bigg{)} \label{ss1},
\end{equation}
\begin{equation}
-\frac{1}{4}g\bigg{(}(f_{\xi})^2+2ff_{\xi\xi}\bigg{)} = \frac{\partial^2S}{\partial\xi^2} \label{ss2}.
\end{equation}

Now, the solutions we seek for $f$ and $g$ should depend only on $\xi$, as otherwise our assumption of the self-similarity of those functions breaks down. Because of the complicated dependence of $S$ on $\Gamma{v_z}$, we see that the solutions will not be self-similar for arbitrary $\Gamma_j$. However, in the ultrarelativistic limit, for which $\Gamma \simeq \Gamma{v_z}$, we see that $S \simeq \ln(\Gamma{v_z})$, and it is apparent that the solutions are indeed self-similar, i.e., the $\zeta$ dependence of equations \eqref{ss1} and \eqref{ss2} drops out. 

As these equations are third order in $f$ and first order in $g$, we require four boundary conditions to solve them numerically. Since $\Gamma{v_z}$ is given by equation \eqref{vzsim2} and it is assumed that $\Gamma{v_z} = \Gamma_jv_j(z/z_0)^{-1/4}$ along the axis, our first boundary condition is given by $f_{\xi}(0) = 1$. For the second, we note that the jet axis is the streamline along which the $y$-component of the velocity is zero. From equation \eqref{vysim}, then, we find $f(0) = 0$ (we will see that $\partial\xi/\partial{z} = 0$ at $\xi = 0$). The third boundary condition is obtained by noting that the $z$-component of the velocity should approach zero as we proceed into the ambient medium, so that, from equation \eqref{vzsim2}, we find $f_{\xi}(\infty) = 0$. Finally, the density of scatterers should approach that of the ambient medium far from the jet center. Equation \eqref{rhosim} then gives $g(\infty) = 1+3/(4\mu)$.

As we noted, the truly self-similar limit of equations \eqref{ss1} and \eqref{ss2} is obtained by setting $S = \ln(\Gamma{v_z})$, but we encounter an issue with this scaling when we consider the boundary conditions on our flow in the ambient medium. In particular, we expect the outflow velocity to approach zero for $\xi \gg 1$; when the velocity becomes subrelativistic, however, $\ln(\Gamma{v_z}) \rightarrow -\infty$, when in actuality the $v_z \ll 1$ limit of equation \eqref{Seq} is $S \simeq \ln(1+\Gamma{v_z}) \simeq \Gamma{v_z} \simeq 0$. Thus, we see that taking the ultrarelativistic limit of equations \eqref{ss1} and \eqref{ss2} will not result in the solutions matching the correct boundary conditions far from the jet center.

To correct this problem, we will allow the function $S$ to take on its full form and treat $\zeta$ as a constant, which will break the self-similarity of our solutions. However, since our boundary conditions do not depend on the value of $\zeta$, the functions $f$ and $g$ themselves should be largely independent of that parameter. We therefore expect the assumption of self-similarity to be upheld in the ultrarelativistic ($\xi \ll 1$) and the non-relativistic ($\xi \gg 1$) limits of our solutions, with small deviations from self-similarity in the trans-relativistic regime of the outflow. 

\begin{figure}[htbp] 
   \centering
   \includegraphics[width=3.5in]{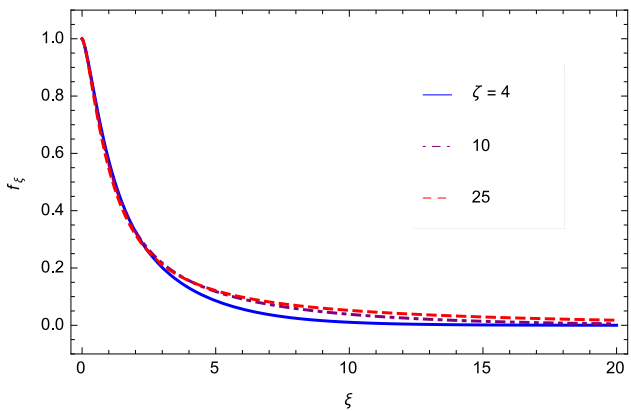} 
   \caption{The behavior of $f_{\xi}$, which is the normalized $z$-component of the four-velocity, with $\xi$ for $\zeta = 4$ (blue, solid curve), $\zeta = 10$ (purple, dot-dashed curve), and $\zeta = 25$ (red, dashed curve). As expected, the curves are all coincident when $\xi \ll 1$ and $\xi \gg 1$, with a non-self-similar transition (one that depends on $\zeta$) in between those two limits.}
   \label{fig:fpsims}
\end{figure}

\begin{figure}[htbp] 
   \centering
   \includegraphics[width=3.5in]{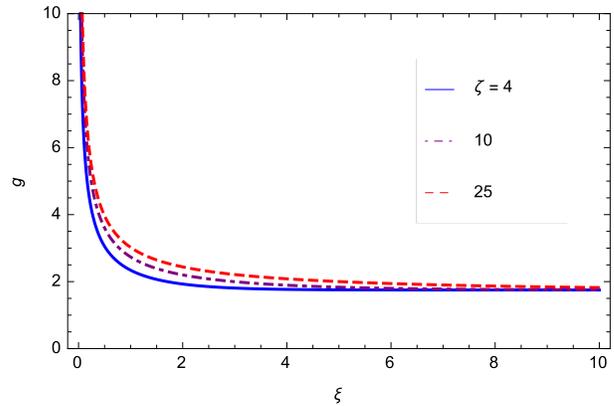} 
   \caption{The function $g$, which is approximately the inverse of the lab-frame density, plotted with respect to $\xi$ for the same set of $\zeta$ chosen in Figure \ref{fig:fpsims}. As was true for $f_{\xi}$, $g$ is approximately self-similar close to and far from the jet, with the deviation from self-similarity in the trans-relativistic region being apparent but small.}
   \label{fig:gsims}
\end{figure}

To exemplify this point, Figures \ref{fig:fpsims} and \ref{fig:gsims} show, respectively, the variation of $f_{\xi}$ -- the normalized $z$-component of the four-velocity -- and $g$ -- approximately the inverse of the observer-frame density -- in terms of the parameter $\xi$ for $\zeta = 4, 10, $ and 25. As we anticipated, the functions for different $\zeta$ are indistinguishable for $\xi \ll 1$ and $\xi \gg 1$, meaning that the self-similarity is nearly exact in those regions. In between those limits, where the flow is trans-relativistic, the deviations from self-similarity are apparent, though they remain small. Keeping the full form of $S$ in equations \eqref{ss1} and \eqref{ss2} therefore preserves well the self-similarity of our solutions and provides a {}{reasonable} interpolation between the relativistic and non-relativistic regions of the flow. 

The solution for $g$ approaches infinity as we near the axis, which can be seen by investigating the small-$\xi$ behavior of equation \eqref{ss2}. Specifically, letting $f \simeq \xi$, $f_{\xi} \simeq 1$, $\Gamma \simeq \zeta$, and $v_z \simeq 1$, equation \eqref{ss2} can be written as

\begin{equation}
\xi{g_{\xi}}+\frac{1}{2}g \simeq -2(f_{\xi\xi,0})^2,
\end{equation}
where $f_{\xi\xi,0}$ is the second derivative of $f$ evaluated at $\xi = 0$. This equation can be integrated, and we find

\begin{equation}
g \simeq \frac{C}{\sqrt{\xi}}-(f_{\xi\xi,0})^2,
\end{equation}
$C$ being a constant of integration. Since it is roughly proportional to $1/\rho'$, the value of $g$ cannot be negative, meaning that $C$ must be greater than zero. Therefore, the asymptotic, $\xi \ll 1$ behavior of $g$ is $g \sim 1/\sqrt{\xi}$. 

In addition to the mass density of scatterers and the velocity, we can also calculate the number density of photons, $N'$, throughout the boundary layer. As we showed in Paper I, the equation of photon number conservation \eqref{fluxeq} becomes, to lowest order in the boundary layer thickness,

\begin{equation}
\nabla_{\mu}\bigg{(}N'U^{\mu}\bigg{)}=\frac{1}{3}\frac{\partial}{\partial{y}}\bigg{(}\frac{1}{\rho'\kappa}\frac{\partial{N'}}{\partial{y}}\bigg{)} \label{Neq1}.
\end{equation}
Performing a few manipulations, this equation becomes

\begin{equation}
U^{\mu}\nabla_{\mu}\bigg{(}\frac{N'}{\rho'}\bigg{)} = \frac{\kappa}{3}\frac{\partial^2N'}{\partial\tau^2} \label{Neq2}.
\end{equation}
As was true for the density of scatterers, we expect the observer frame number density of photons to vary approximately self-similarly, which will be true if we let 

\begin{equation}
\frac{N'}{\rho'} = \frac{N'_0}{\rho'_0}h(\xi) \label{Nselfsim},
\end{equation}
where $N'_0$ is the number density of photons in the ambient medium and $h$ is a dimensionless function. Inserting this ansatz into equation \eqref{Neq2} gives

\begin{equation}
-\frac{1}{3}f\,h_{\xi} = \zeta\frac{\partial^2}{\partial\xi^2}\bigg{(}\frac{h}{\Gamma{g}-\frac{3}{4\mu}}\bigg{)} \label{heq}.
\end{equation}
For the boundary conditions on $h$, we first require that the number density of photons approach that of the ambient medium in the $\xi \gg 1$ limit. Equation \eqref{Nselfsim} then gives $h(\infty) = 1$. For the second condition, return to equation \eqref{Neq1}, integrate both sides from $y = -\infty$ to $\infty$, and require that the derivative of $N$ vanish in both of those limits. Doing so yields

\begin{equation}
\frac{\partial}{\partial{z}}\int_{-\infty}^{\infty}N'\Gamma{v_z}dy = -N'\Gamma{v_y}\bigg{|}_{-\infty}^{\infty} \label{hbc}.
\end{equation}
The right-hand side can be determined by returning to the continuity equation, integrating from $y = -\infty$ to $\infty$ and performing a few manipulations to show that $\Gamma{v_{\infty}}\big{|}_{-\infty}^{\infty} = -f_{\infty}$, where $f_{\infty}$ is the function $f$ evaluated at infinity. Using the definition of $N'$ in terms of $h$, we find that equation \eqref{hbc} becomes

\begin{equation}
\int_{-\infty}^{\infty}h\,f_{\xi}\,d\xi = 2\,f_{\infty},
\end{equation}
which serves as our second boundary condition on $h$. This integral states that the increase in the number flux of photons occurs at a rate provided by the influx of material at infinity. 

The equations derived in this section were all written in terms of the variable $\xi$, which is itself a function of $g$ via equations \eqref{taueq} and \eqref{xieq}. To write the solutions in terms of the physical parameter $y$, we can return to equation \eqref{taueq}, differentiate both sides with respect to $y$, rearrange the resulting equation and integrate to yield

\begin{equation}
\int_0^{\xi}\bigg{(}\Gamma{g}-\frac{3}{4\mu}\bigg{)}d\tilde\xi = y\sqrt{\frac{9}{2}\frac{\Gamma_j^2v_j^2\rho'_0\kappa}{z_0}}\bigg{(}\frac{z}{z_0}\bigg{)}^{-3/4} \equiv \alpha \label{xiofalpha},
\end{equation}
where $\tilde\xi$ is a dummy variable of integration. Once we calculate the functions $g(\tilde\xi)$ and $f_{\xi}(\tilde\xi)$, this relation can be integrated and solved numerically to yield $\xi(\alpha)$. This expression also shows that

\begin{equation}
\bigg{(}\Gamma{g}-\frac{3}{4\mu}\bigg{)}\frac{\partial\xi}{\partial{z}} = \frac{1}{4\,z}\bigg{(}\int_0^{\xi}\frac{\zeta^2(f_{\xi})^2g}{\sqrt{1+\zeta^2(f_{\xi})^2}}d\tilde\xi-3\,\alpha\bigg{)} \label{dxidz},
\end{equation}
which we can use in equation \eqref{vysim} to give

\begin{multline}
\rho'\kappa\Gamma{v_y} = \\ 
\frac{1}{4}\sqrt{\frac{2}{9}\frac{\rho'_0\kappa}{z}}\bigg{(}\frac{f_{\xi}}{\Gamma{g}-\frac{3}{4\mu}}\bigg{(}3\alpha-\int_0^{\xi}\frac{\zeta^2(f_{\xi})^2g}{\sqrt{1+\zeta^2(f_{\xi})^2}}d\tilde\xi\bigg{)}-2f\bigg{)} \label{vy2}
\end{multline}
Equation \eqref{dxidz} also confirms that $\partial\xi/\partial{z} = 0$ when $\xi = 0$, which we used in order to determine the boundary condition $f(0) = 0$. 

\subsection{Solutions}
In this section we plot solutions for the outflow velocity, the density of scatterers and the density of photons for various values of $\mu$ and $\zeta$. As was mentioned in the previous subsection, the physical self-similar variable against which we would like to plot our solutions is given by $\alpha$. However, as is apparent from equation \eqref{xiofalpha}, the definition of $\alpha$ depends on $\zeta$. Therefore, if we are comparing, for example, the outflow velocity of two systems with differing $\zeta$, we must incorporate the $\zeta$ dependence in $\alpha$ so that the range of physical space that we consider for each solution is the same. For this reason, in this section we will plot our solutions as functions of the variable

\begin{equation}
\tilde\alpha \equiv \frac{\alpha}{\zeta} = y\sqrt{\frac{9}{2}\frac{\rho'_0\kappa}{z_0}}\bigg{(}\frac{z}{z_0}\bigg{)}^{-1/2} \label{alphatilde}.
\end{equation}

\begin{figure}[htbp] 
   \centering
   \includegraphics[width=3.5in]{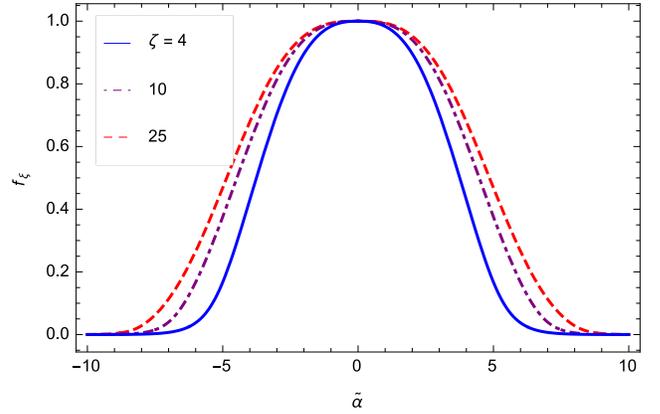} 
   \caption{The normalized $z$-component of the four-velocity ($f_{\xi}$) for $\mu = 1$ and $\zeta = 4$, 10, and 25 (the solid, blue curve, the dot-dashed, purple curve, and the dashed, red curve, respectively), which, for $z \simeq z_0$, correspond to $\Gamma_j = 4, 10, $ and 25. We see that the width of the boundary layer is nearly unchanged as we alter the value of $\zeta$. }
   \label{fig:fpplotsgamjet}
\end{figure}

\begin{figure}[htbp] 
   \centering
   \includegraphics[width=3.5in]{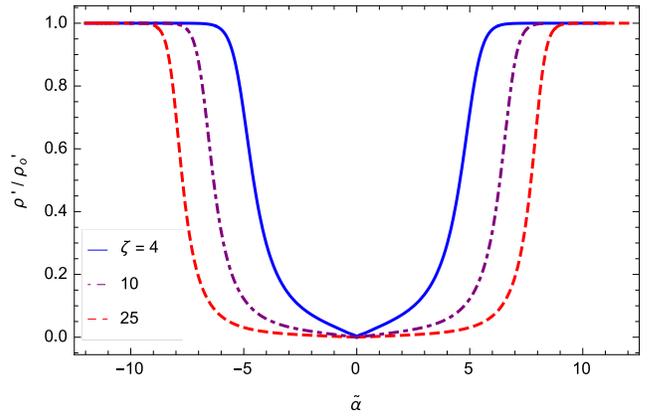} 
   \caption{The normalized fluid-frame density of scatterers for the same set of parameters used in Figure \ref{fig:fpplotsgamjet}. For all solutions the number density of scatterers approaches zero as we near the center of the jet. We see that the average number density of scatterers within the boundary layer is lower for larger Lorentz factors.}
   \label{fig:rhoplotsgamjet}
\end{figure}

\begin{figure}[htbp] 
   \centering
   \includegraphics[width=3.5in]{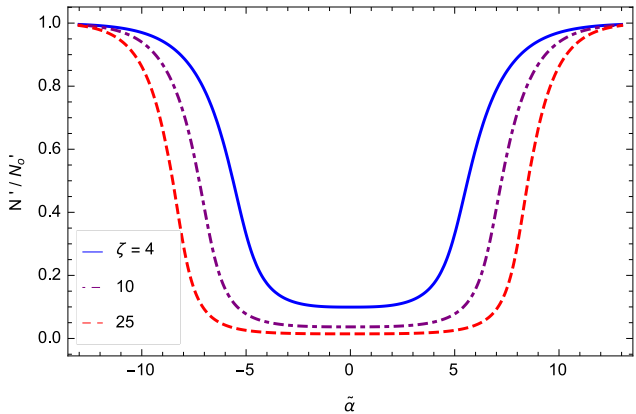} 
   \caption{The normalized number density of photons for the same set of $\zeta$ used in Figure \ref{fig:fpplotsgamjet}. The photon number density closely follows that of the scatterers.}
   \label{fig:hplotsgamjet}
\end{figure}

\begin{figure}[htbp] 
   \centering
   \includegraphics[width=3.5in]{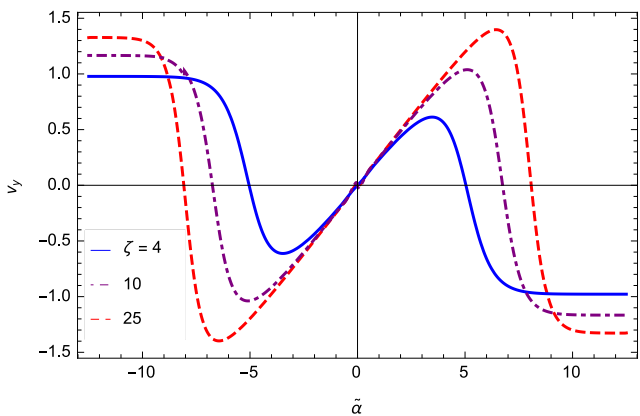} 
   \caption{The $y$-component of the three-velocity normalized by $\sqrt{2/(9\rho'_0\kappa{z_0})}$ (see equation \eqref{vy2}) for the same set of parameters used in Figure \ref{fig:fpplotsgamjet}. For positive $\tilde\alpha$, each solution is initially positive and reaches a relative maximum before approaching a negative constant, which shows that the flow expands outwards near the center of the jet and entrains ambient material far from the axis. }
   \label{fig:vyplotsgamjet}
\end{figure}

Figure \ref{fig:fpplotsgamjet} shows the solution for $f_{\xi}$, the normalized $z$-component of the four-velocity, for $\zeta = 4$, 10, and 25. Since $\zeta = \Gamma_jv_j(z/z_0)^{-1/4}$, these values of $\zeta$ scale approximately linearly with the Lorentz factor until $z \gg z_0$. The outflow velocity is maximized at the origin and decays as we move farther into the ambient medium. We see that the width of the boundary layer, loosely defined as the value of $\tilde\alpha$ at which $f_{\xi}$ is some fraction of its central value, is nearly unchanged as we modify $\zeta$. The average value of the normalized $z$-component of the velocity is also slightly larger for larger $\zeta$.

Figure \ref{fig:rhoplotsgamjet} demonstrates how the normalized, fluid-frame mass density of scatterers varies as we traverse the boundary layer for the same set of parameters chosen in Figure \ref{fig:fpplotsgamjet}. Since $g$ approaches infinity as we near the origin, the mass density of scatterers, related to $g$ by equation \eqref{rhosim}, equals zero at the origin for all of the solutions, meaning that the center of the jet is evacuated of massive particles. The average comoving density of scatterers across the boundary layer is also lower for larger $\zeta$, in accordance with equation \eqref{rhosim}. In Figure \ref{fig:hplotsgamjet} we plot the normalized number density of photons for the same set of $\zeta$. It is evident that the density of photons closely follows the density of scatterers throughout the boundary layer. We see, however, that the density of photons stays above and below the density of scatterers as we move toward and away from the center of the jet, respectively. The photon density also remains finite at the center of the jet. 

The $y$-component of the three-velocity normalized by $\sqrt{2/(9\rho'_0\kappa{z_0})} \sim \delta$ is illustrated in Figure \ref{fig:vyplotsgamjet} for the same set of $\zeta$ used in Figure \ref{fig:fpplotsgamjet}. For $\xi > 0$, each solution initially has a positive ${v_y}$, which shows that the jet material expands away from the axis. The $y$-component of the three-velocity then reaches a relative maximum, one which increases slightly for larger $\Gamma_j$, before approaching a negative, constant value. This behavior is then inverted for negative $\xi$. Because the transverse velocity approaches a negative constant for $\xi >0$ and a positive constant for $\xi < 0$, we see that the jet entrains material from the ambient medium. 

\begin{figure}[htbp] 
   \centering
   \includegraphics[width=3.5in]{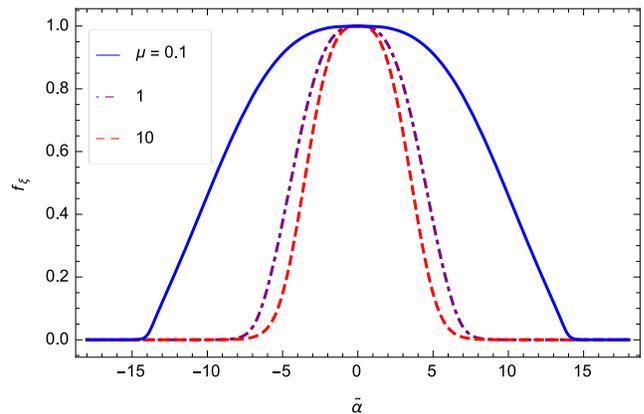} 
   \caption{The function $f_{\xi}$, which is the normalized $z$-component of the four-velocity, for $\zeta = 10$ and $\mu = e'/\rho'_0 = 0.1$, 1, and 10, which correspond to the blue, solid curve, the purple, dot-dashed curve, and the red, dashed curve, respectively. Increasing the value of $\mu$, we see, has little effect on the solution, while decreasing $\mu$ drastically widens the boundary layer.}
   \label{fig:fpplotsmujet}
\end{figure}

\begin{figure}[htbp] 
   \centering
   \includegraphics[width=3.5in]{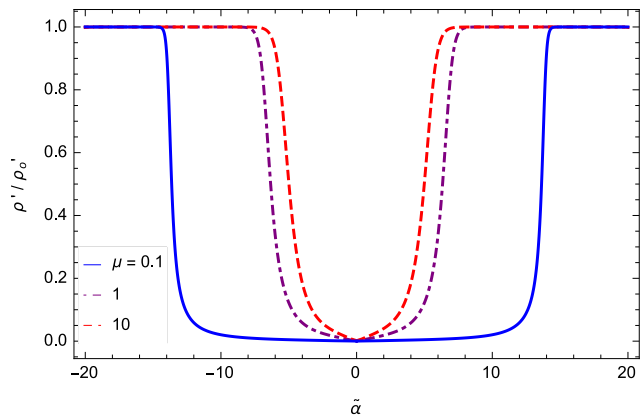} 
   \caption{The normalized density of scatterers for the same values of $\mu$ chosen in Figure \ref{fig:fpplotsmujet}. The mean value of the density decreases as $\mu$ decreases.}
   \label{fig:rhoplotsmujet}
\end{figure}

\begin{figure}[htbp] 
   \centering
   \includegraphics[width=3.5in]{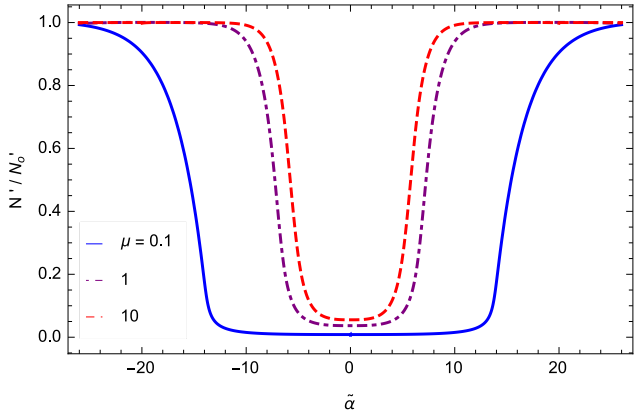} 
   \caption{The normalized number density of photons for the same set of $\mu$ chosen in Figure \ref{fig:fpplotsmujet}. This figure demonstrates, as we saw in Figure \ref{fig:hplotsgamjet}, that the density of photons tracks that of the scatterers.}
   \label{fig:hplotsmujet}
\end{figure}

\begin{figure}[htbp] 
   \centering
   \includegraphics[width=3.5in]{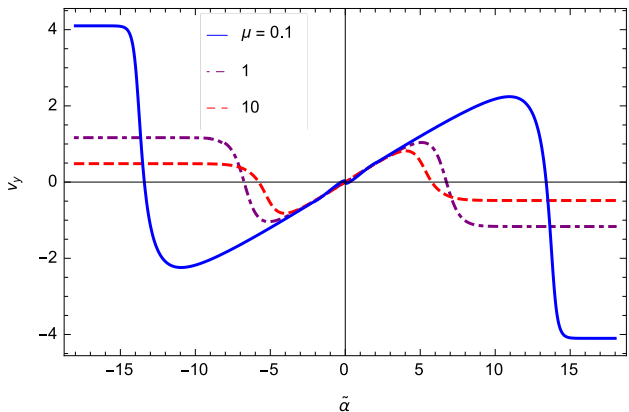} 
   \caption{The $y$-component of the three-velocity, normalized by $\sqrt{2/(9\rho'_0\kappa{z_0})} \sim \delta$, for the same set of $\mu$ chosen in Figure \ref{fig:fpplotsmujet}. We see that the relative maximum increases for smaller $\mu$.}
   \label{fig:vyplotsmujet}
\end{figure}

Figures \ref{fig:fpplotsmujet} -- \ref{fig:vyplotsmujet} illustrate how our solutions depend on $\mu$. As is apparent, changing the value of $\mu$ does not drastically alter the qualitative aspects of the functions. We do see, however, that {}{decreasing $\mu$ from 1 to 0.1 results in} a large increase in the boundary layer thickness; on the contrary, {}{changing $\mu$ from 1 to 10} results in only a slight narrowing of its thickness. It is also evident that a smaller $\mu$ {}{compared to 1} results in a lower average value of the density throughout the boundary layer and a larger peak in the transverse velocity ${v_y}$. 

\section{Discussion}
Figures \ref{fig:fpplotsgamjet} -- \ref{fig:vyplotsgamjet} demonstrate that the width of the boundary layer is nearly independent of $\zeta$, which is the result of the scaling of our boundary layer thickness $\delta$. Specifically, note from equation \eqref{deltaeq} that $\delta$ is given by $\delta \sim 1/\sqrt{\rho'\,\Gamma_j}$. Since our ansatz posited that $\rho'\Gamma\simeq g$, with $g$ a dimensionless function of order unity, and $\Gamma \simeq \Gamma_j \simeq \zeta$  for $z \sim z_0$, the boundary layer thickness is roughly independent of $\zeta$.

Figures \ref{fig:fpplotsmujet} -- \ref{fig:vyplotsmujet} illustrate that a value of $\mu = 0.1$ causes the boundary layer thickness to increase dramatically {}{compared to $\mu = 1$}, while {}{setting} $\mu = 10$ causes only a slight narrowing of the width {}{compared to $\mu = 1$}. This dependence is due to the fact that the viscous heating, which increases the specific entropy by decreasing the density of scatterers (at fixed pressure), is most efficient when the flow is compressible, as is evident from the gas energy equation \eqref{gasen1}. The fluid only becomes compressible, however, when the flow is supersonic, and we can show \citep{cou14b} that the sound speed of a radiation-dominated gas is

\begin{equation}
c_s = \frac{2}{3}\sqrt{\frac{\mu}{1+4\mu/3}} \label{cseq}.
\end{equation}
When $\mu \ll 1$, the sound speed reduces to $c_s \sim \sqrt{\mu}$, and the location at which the flow becomes transsonic extends farther into the ambient medium, widening the boundary layer. Conversely, when $\mu \gg 1$, the sound speed approaches a constant $\simeq 1/\sqrt{3}$, which results in only a slight narrowing of the boundary layer. 

Equation \eqref{ymom1} shows that the energy density of the radiation, and hence the pressure, is constant across the jet, which is a statement of the causal connectedness of the boundary layer. In order for this equation to remain valid, then, we require that the transverse sound crossing time over the boundary layer thickness $\delta{y}$ be less than the time it takes the fluid to traverse the distance $\delta{z}$. Since the transverse sound speed is given $c_{\perp} = c_s/\Gamma$, where $c_s$ is given by equation \eqref{cseq}, this requirement yields the inequality $\delta \lesssim c_s/(\Gamma_j{v_j})$. Once this inequality is no longer satisfied, equation \eqref{ymom1} does not hold, and we must include more terms in equations \eqref{ymom1} -- \eqref{gasen1} to account for the gradients in the energy density of the radiation. 

As we noted in Section 3, the density approaches zero as we near the axis of the jet. Physically, this effect arises from the fact that, for $z \ll z_0$, the Lorentz factor grows unbounded and the boundary layer thickness goes to zero. Therefore, the center of the jet originates along a curve of infinite shear and, consequently, infinite entropy. Since the specific entropy scales as $s' \propto e'/(\rho')^{4/3}$ and $e'$ is constant across the boundary layer, we see that the density of scatterers must equal zero at the center of the jet.

From Figures \ref{fig:vyplotsgamjet} and \ref{fig:vyplotsmujet}, we see that the $y$-component of the velocity initially causes the outflow to expand into its surroundings. When $|\xi|$ becomes large, however, the directionality of ${v_y}$ reverses toward the jet. The outflow therefore entrains material from the ambient medium not only by expanding in the transverse direction, but also by dragging material in from the environment.

{}{Interestingly, the $y$-component of the velocity approaches a non-zero, constant value as we move into the external medium. This behavior arises from the fact that the jet is removing material from the system at a rate $\dot{M} \sim \rho'\Gamma{v_z}dy \sim z^{1/2}$, showing that amount of mass excavated from the envelope increases as we move along $z$. Therefore, in order to maintain a steady-state, we require a constant influx of material at infinity that can resupply the amount lost due to the jet. This interpretation is substantiated by integrating the continuity equation \eqref{masscont1} from $y = -\infty$ to $y = \infty$, which shows that the $y$-component of the velocity at infinity scales as $v_{y,{\infty}} \sim -f_{\infty}/2$. The factor of $1/2$ arises from the fact that the mass loss rate scales as $\dot{M} \sim \rho'\Gamma{v_z}dy \propto z^{1/2}$; therefore, if one could decrease the mass loss rate to one that was constant in $z$, then the $y$-component of the velocity would vanish at infinity. Likewise, if one could create a scenario in which $\dot{M}$ scaled as a \emph{negative} power of $z$, then the value of $v_y$ would maintain an efflux of material at infinity to keep the system from amassing inertia towards the center of the jet (thus violating the steady-state assumption).}

The free-streaming jet solutions analyzed in this paper are only valid when the Lorentz factor is relativistic, i.e., when $\zeta > 1$. When $\zeta \ll 1$, $\Gamma \simeq 1$, $\rho' \simeq \rho'_0$, and we can show that conservation of momentum along the $z$-axis results in the scaling $v_j \sim z^{-1/3}$, $\delta \sim z^{2/3}$, which agrees with the results of the incompressible, non-relativistic theory {}{(in Chapter 10, Section 12 of \citealt{kun08}, see their discussion at the top of page 383)}. Therefore, once $z \gtrsim \Gamma_j^4\,z_0$, the outflow will undergo a non-self-similar transition from the solution presented here to its non-relativistic counterpart. 

The free-streaming jet model has observational consequences. For example, the decrease in the density along the jet axis means that observers looking down the barrel of the jet see farther into the outflow. Because $\Gamma \propto z^{-1/4}$ and $e = \Gamma^2e'$, where $e$ is the lab-frame radiation energy density, this means that those observers see a higher energy density of photons. Also, those same observers, using the fact that the lab-frame radiation number density is $N \sim \Gamma{N'}$, see an energy per photon of $e/N \sim \Gamma{e'}/N'$. Therefore, not only do the on-axis observers see a more Lorentz-boosted spectrum because they can see deeper into the outflow, their spectrum is also hardened from the fact that $N'$ is minimized along the axis of the jet {}{(where $y =0$ in the figures of Section 3)}. 

\subsection{Comparative notes between the two-stream model (Paper I) and the free-streaming jet model}

{}{Paper I analyzed how the presence of a radiation pressure-supported envelope affected the propagation of a radiation-dominated, relativistic jet through the two-stream approximation. This approximation treats the jet and its surroundings as semi-infinite, separate fluids, a contact discontinuity serving as the surface of separation between the two. In contrast, the free-streaming jet solution presented here treats the entire system -- the jet and its surroundings -- as a single fluid. There is thus no formal distinction between ``jet material'' and ``ambient material,'' meaning that no contact discontinuity exists in the system.}

{}{By comparing Figure 1 of Paper I and Figure \ref{fig:fpplotsgamjet} of the previous section, we see that the $z$-component of the four-velocity at a fixed $z$ behaves similarly between the two models. Namely, the solution starts at some ``jet'' velocity, which corresponds to $y = \infty$ in the two-stream model and to $y = 0$ in the jet model, and smoothly transitions to a velocity of approximately zero over the extent of a few boundary layer thicknesses. However, the full spatial dependence of the velocity, one that includes variation in the $z$-direction, differs drastically between the two models: while the two-stream jet maintains a constant $\Gamma_j$ along $z$, we found here that the $z$-component of the four-velocity scales as $\Gamma{v_z} \propto z^{-1/4}$, implying that the overall velocity of the jet slows as we look farther down $z$. This behavior arises from the fact that the free-streaming jet can entrain material, this entrainment causing an increase in the inertia contained in the outflow and a resultant decrease in its velocity.}

{}{The general behavior of the number densities of scatterers and photons between the models is also similar, which can be understood by comparing Figures 2 and 3 of Paper I to Figures \ref{fig:rhoplotsgamjet} and \ref{fig:hplotsgamjet} of Section 3, respectively. In particular, both densities decrease within the boundary layer separating the outflow and its surroundings, asymptotically approaching their jet values as $y\rightarrow \infty$ in the two-stream model and as $y\rightarrow 0$ in the free-streaming jet model. Likewise, each approaches its ambient value as $y \rightarrow -\infty$ in the two-stream model and as $y \rightarrow \pm \infty$ in the free-streaming jet model. Although the gross properties of both are similar, one striking difference arises, however, in the behavior of the scatterer density: in the two-stream model, the existence of the contact discontinuity causes the function $g$, and hence the density $\rho'$, to vanish within the boundary layer (it vanishes specifically at $y = 0$ -- the location of the contact discontinuity). Conversely, $\rho'$ remains non-zero throughout the boundary layer that connects the jet and its surroundings in the model presented here, and only as we near the center of the jet does the density of scatterers go to zero. The presence of a contact discontinuity thus results in the likely non-physical vanishing of the massive particles within the boundary layer.}

{}{Finally, the scaling of the boundary layer thickness itself differs between the two models. In the two-stream case, we found that $\delta \propto z^{1/2}$ (see equation (15) of paper I). Therefore, as one looks farther down the $z$-axis, the boundary layer that develops between the jet and the ambient environment extends into both media at a rate proportional to $z^{1/2}$. Contrarily, equation \eqref{xiofalpha} shows that the free-streaming jet boundary layer expands into its surroundings as $\propto z^{3/4}$ (though this transitions to $z^{2/3}$ in the non-relativistic limit; see the discussion at the top of page 383 in Chapter 10, Section 12 of \citealt{kun08}). The free-streaming jet boundary layer thus expands more rapidly than does the two-stream solution.}

\section{Summary and conclusions}
Employing the equations of radiation hydrodynamics in the viscous limit, which are applicable as long as changes in fluid quantities are small over the mean free path of a photon, we analyzed the dynamics of a relativistic, free-streaming jet under the boundary layer approximation. This approximation, which should be upheld in jets with transverse optical depths substantially greater than one, states that variations in the properties of the outflow are confined to a thin layer of width $\delta$, and it allowed us to transform the full set of equations \eqref{radhydroco} -- \eqref{masscont} into a set of greatly simplified boundary layer equations \eqref{ymom1} -- \eqref{masscont1}. 

Perhaps the biggest difference between the two-stream solutions, presented and analyzed in paper I, and the free-streaming jet solutions presented here is in the distinction between the jet and the ambient medium. In the former, the two are considered as distinct, interacting entities, which allows one to specify separately their asymptotic properties. The latter approach, on the other hand, considers the whole configuration as a single fluid.

Because one has more freedom in specifying the properties of the outflow, the two-stream solution has the added benefit of being able to treat scenarios in which the properties of the jet and the ambient medium differ significantly. However, maintaining the distinction between the two media necessitates the existence of a surface of contact between the two across which no fluid can flow, meaning that the jet cannot entrain ambient material. Furthermore, as was demonstrated in Paper I, this boundary condition results in the density formally vanishing at the interface, which is likely non-physical.

By treating the outflow and the environment as one fluid, we demonstrated that entrainment does occur in the radiation-viscous, free-streaming jet solution, which causes the $z$-component of the four-velocity of the jet to slow as $\sim z^{-1/4}$ (this power-law, however, may differ if the adopted symmetry is azimuthal as opposed to planar; see below). These solutions also show that the comoving densities of scatterers and photons decrease dramatically within the boundary layer, with the density of scatterers approaching zero as one nears the center of the jet. Therefore, because observers that look ``down the barrel of the jet'' can see farther into the outflow, they see a more Lorentz-boosted energy density than those that view the outflow off-axis. They also observe a higher energy per photon, given by $e/N \simeq e'\Gamma_o/N'$, both because $N'$ is minimized along the axis and because $\Gamma_o$, the observed Lorentz factor, is larger. Such features are in qualitative agreement with the event \emph{Swift} J1644+57, where such an observer orientation is invoked to explain the X-ray emission \citep{zau11}. 

{}{In addition to super-Eddington TDEs and long GRBs (as well as the relatively new class of ``ultra-long'' GRBs; \citealt{lev14}) -- the two applications considered in the Introduction -- the free-streaming jet model could be applied to other sources. As mentioned in Paper I, this model may also be relevant to microquasars \citep{fen04} (particularly those that fall in the class of ULX's; \citealt{kin01}) such as the object SS 433 (\citealt{fab04, beg06a}; see \citet{ara93} for an application of the two-stream model to this source). Additionally, a jetted quasi-star -- a protogalactic gas cloud supported by a supercritically accreting black hole \citep{beg06b, beg08, cze12} -- provides another situation in which a radiation-dominated jet propagates alongside a radiation pressure-supported envelope.}

{}{The free-streaming jet solution presented here is limited to describing plane-parallel, two-dimensional systems. One consequence of this assumption is that the entrainment of material, which slows down the jet as $\Gamma \propto z^{-1/4}$, happens effectively in one dimension. If, on the other hand, one imposed azimuthal symmetry and described the system in terms of spherical coordinates $(r,\,\theta,\,\phi)$, which is likely more relevant for realistic jetted systems, the entrainment would occur in two dimensions. Other conditions being equal, the slowing of the jet along the axis would then be more pronounced, i.e., the Lorentz factor would scale as $\Gamma \propto r^{-p}$ with $p > 1/4$.}

{}{The other main assumption of the free-streaming jet model presented here is that the energy density of the ambient medium is independent of $z$, i.e., $e'(z) = e'_0$. If a pressure gradient were present, this force would tend to accelerate (or, in principle, decelerate) the jet material, offsetting the power-law scaling $\Gamma \propto z^{-1/4}$. In fact, one can imagine that if the pressure gradient were strong enough, it could indeed reverse the overall slowing of the jet and cause the outflow to accelerate. If the energy density scaled as $e' = e'_0(z/z_0)^{-q}$, which is relevant for super-Eddington TDEs as long as the constant $q$ satisfies $q >3/2$ \citep{cou14a}, this situation could be actualized near the launch point of the jet.}

In an ensuing paper, we plan to compare more quantitatively the predictions made by the free-streaming jet model presented here and the observations of \emph{Swift} J1644+57. We will also extend our analysis to incorporate a spherical geometry -- as azimuthal symmetry is almost certainly more relevant for this and other systems than the planar symmetry adopted here -- as well as radially-dependent ambient pressure and density profiles.

\acknowledgements
This work was supported in part by NASA Astrophysics Theory Program grant NNX14AB37G, NSF grant AST-1411879, and NASA's Fermi Guest Investigator Program. 

{}

\end{document}